\documentclass[prd,aps,nofootinbib,onecolumn,showpacs,10pt]{revtex4}
\usepackage{graphicx,epsfig}

\usepackage{epsf}
\usepackage{graphicx}

\begin{document}

\title{  \bf Semileptonic $B_{c}$
to P-Wave Charmonia $(X_{c0}, X_{c1}, h_{c})$ Transitions within
QCD Sum Rules }
\author{ K. Azizi$^{1,\dag}$ ,  H. Sundu$^{2,\ddag}$,  M. Bayar$^{2,*}$
\\$^1$Department of Physics, Middle East Technical University, 06531 Ankara, Turkey\\
$^2$Physics Department, Kocaeli University, 41380 Izmit, Turkey\\
$^\dag$e-mail:e146342@metu.edu.tr \\
$^\ddag$e-mail:hayriye.sundu@kocaeli.edu.tr \\
$^*$e-mail: melahat.bayar@kocaeli.edu.tr}

\begin{abstract}

The form factors of the semileptonic $B_c\rightarrow S(AV)\ell\nu$
$(\ell=\tau, \mu, e)$ transitions, where S and AV denote the scalar
$X_{c0}$ and axial vector $(X_{c1},\,\,h_{c})$ mesons, are
calculated within the framework of the three-point QCD sum rules. The heavy quark
effective theory limit of the form factors are also obtained and compared with the values of the original transition form factors. The results of  form factors are
used to estimate the total decay widths and branching ratios of
these transitions. A comparison of our results on branching ratios with the predictions
of other approaches is also presented.

\end{abstract}

 \maketitle

\section{Introduction}
The $B_c$ meson is the only known meson composed of two heavy quarks
of different flavor and charge; a charm quark and a bottom
antiquark. It were discovered by the collider detector at Fermilab
(CDF Collaboration) in $p\overline{p}$ collision via the decay mode
$B_c\rightarrow J/\psi l^{\pm}\nu$ at $\sqrt{s}=1,8~TeV$ \cite{1}.
Discovery of the $B_c$ meson has demonstrated the possibility of the
experimental study of the charm-beauty system and has created
considerable interest in its spectroscopy \cite{2, 3, 4, 5, 6}. When
the large hadron collider (LHC)  runs, a  plenty number of $B_c$ events, which are expected to be about
$10^8 \sim 10^{10}$ per year with the the luminosity values of
$L=10^{34} cm^{-2}s^{-1}$ and $\sqrt{s}=14~TeV$, will be produced
 \cite{7, 8}. Therefore, not only experimental but also theoretical
study on $B_c$ mesons will be of great interests in many respects.

Among the B mesons, the $B_c$  carries a distinctive signature and
has reached great interest recently for the following reasons:
Firstly, the $B_c$ meson decay channels are expected to be very rich
in comparison with other B mesons, so investigation of such type of
decays can be used in the calculation of the
Cabibbo-Kabayashi-Maskawa (CKM) matrix elements, leptonic decay
constant as well as the origin of the CP and T violation. Secondly,
the $B_c$ meson, because of containing the heavy quarks, provides
more accuracy and confidence in the understanding of the QCD
dynamics.

The $B_c$ meson can decay via the $b\rightarrow u,d,s,c$ and also
the $c\rightarrow u,d,s$ transitions. Among those transitions at
quark level, the tree level $ b \rightarrow c$ transition, governs
the $B_c$ to P-wave charmonia, plays a significant role, because
this  is the most dominant transition. In the literature,
there are several studies on the $B_c$ mesons in different models.
Some possible $B_c$ meson decays such as $B_{c}\rightarrow l
\overline{\nu}\gamma$, $B_{c}\rightarrow \rho^{+}\gamma$,
$B_{c}\rightarrow K^{\ast+}\gamma$ and $B_{c}\rightarrow
B_{u}^{\ast}l^{+}l^{-}$, $B_{c}\rightarrow B_{u}^{\ast}\gamma $,
$B_{c}^{-}\rightarrow D^{*0}\ell\nu$, $B_c \rightarrow P(D,D_s)
l^{+}l^{-}/\nu\bar{\nu}$, $B_c \to D_{s,d}^{*} l^+ l^-$, $B_c
\rightarrow X\nu\bar{\nu}$ and $B_c \rightarrow D_{s}^{*}\gamma$ have been studied in the
frame of light-cone QCD and three-point QCD sum rules
\cite{Aliev1,Aliev2,Aliev3,Alievsp,Kazým1,Kazým2,Kazým3,Kazým4,Kazým5}. The weak productions of new charmonium in semileptonic decays of $B_c$ were also studied in the framework of light cone QCD sum rules in \cite{yuming}. In \cite{Ivanov}, a larger set of exclusive nonleptonic and
semileptonic decays of the $B_c$ meson were investigated in the
relativistic constituent quark model. Weak decays of the $B_c$ meson to charmonium and D mesons in the relativistic quark model have been discussed in \cite{galkin1,galkin2}. Moreover, the $B_c\rightarrow
(D^{\ast}, D^{\ast}_{s})\nu \overline{\nu}$ transitions were also studied
within the relativistic constituent quark model in \cite{Faessler}.

Present work is devoted to the study of the $B_c\rightarrow S(AV)
\ell \nu$. The long distance dynamics of such transitions can be
parameterized in terms of some form factors  which play
fundamental role in analyzing  such transitions. For evaluation of
the form factors, the QCD sum rules as nonperturbative approach
based on the fundamental QCD Lagrangian is used. The obtained
results for the form factors are used to estimate the total decay
rate and branching fractions for the related transitions. The heavy quark effective theory (HQET) limit of the form factors are also calculated and compared with their values. In these transitions, the main contribution comes from the perturbative part since the heavy quark condensates are suppressed by inverse of the heavy quark masses and can be safely omitted and two-gluon condensate contributions are  very small and we will ignore them.  Note that,
the $B_c$ to P-wave charmonia transitions have also been
investigated in the framework of covariant light-front quark model
(CLQM), the renormalization group method (RGM), relativistic
constituent quark model (RCQM) and nonrelativistic constituent quark
model (NRCQM) in \cite{Wang, Chang:2001pm, Ivanov:2006ni,
Ivanov:2005fd, Hernandez:2006gt}. For more about those transitions
see also
\cite{BcXchc1,BcXchc2,BcXchc3,BcXchc4,BcXchc5,BcXchc6,BcXchc7,BcXchc8}.

The outline of the paper  is as follows: In Section II, the some
rules for the transition form factors relevant to the
$B_c\rightarrow S(AV) \ell \nu$ decays are obtained. Section III encompasses the calculation of the HQET limit of the form factors and, section IV is  devoted to   the numerical analysis of the
form factors and their HQET limits, decay rates, branching ratios, conclusion and
comparison of our results with the other approaches.

\section{Sum rules for the $B_{c}\rightarrow S(AV)\ell\nu$ transition form factors }
The $B_{c}\rightarrow X_{c0}\,\,(X_{c1},\,\,h_{c})\ell\nu$ decays
proceed via the $b\rightarrow c$ transition at the quark level. The
effective Hamiltonian responsible for  these transitions can be
written as:

\begin{equation}\label{lelement}
H_{eff}=\frac{G_{F}}{\sqrt{2}} V_{cb}~\overline{\nu}
~\gamma_{\mu}(1-\gamma_{5})l~\overline{c}
~\gamma_{\mu}(1-\gamma_{5}) b .
\end{equation}
We need to sandwich Eq. (\ref{lelement}) between initial and final
meson states in order to obtain the matrix elements of
$B_{c}\rightarrow S(AV)\ell\nu$. Hence, the amplitude of this decay
is written as follows:
\begin{equation}\label{2au}
M=\frac{G_{F}}{\sqrt{2}} V_{cb}~\overline{\nu}
~\gamma_{\mu}(1-\gamma_{5})l<S(AV)(p')\mid~\overline{c}
~\gamma_{\mu}(1-\gamma_{5}) b\mid B_{c}(p)>.
\end{equation}
It is necessary to calculate the matrix elements
$<S(AV)(p')\mid\overline{c}\gamma_{\mu}(1-\gamma_{5}) b\mid
B_{c}(p)>$ appearing in Eq. (\ref{2au}).
 In the S case in final state, the only
axial-vector part of the transition current,
$~\overline{c}~\gamma_{\mu}(1-\gamma_{5}) b~$, contribute to the
matrix element stated above. However, in the AV case, both vector
and axial-vector parts have contributions. Considering the parity
and Lorentz invariances, the aforementioned matrix element can be
parameterized in terms of the form factors in the following way:
\begin{equation}\label{3au}
<S(p')\mid\overline{c}\gamma_{\mu}\gamma_5 b\mid
B_c(p)>=f_{1}(q^{2})P_{\mu}+f_{2}(q^{2})q_{\mu},
\end{equation}
\begin{equation}\label{32au}
<AV(p',\varepsilon)\mid\overline{c}\gamma_{\mu}\gamma_{5} b\mid
B_c(p)>=i\frac{f_{V}(q^2)}{(m_{B_{c}}+m_{X_{c1}})}\varepsilon_{\mu\nu\alpha\beta}
\varepsilon^{\ast\nu}p^\alpha p'^\beta,
\end{equation}
\begin{eqnarray}\label{42au}
< AV(p',\varepsilon)\mid\overline{c}\gamma_{\mu}  b\mid B_{c}(p)>
&=&i\left[f_{0}(q^2)(m_{B_{c}} +m_{X_{c1}})\varepsilon_{\mu}^{\ast}
 -
\frac{f_{+}(q^2)}{(m_{B_{c}}+m_{X_{c1}})}(\varepsilon^{\ast}p)P_{\mu}
-
\frac{f_-(q^2)}{(m_{B_{c}}+m_{X_{c1}})}(\varepsilon^{\ast}p)q_{\mu}\right],
\end{eqnarray}
where $f_{1}(q^2)$, $f_{2}(q^2)$,  $f_{V}(q^2)$, $f_{0}(q^2)$,
$f_{+}(q^2)$ and $f_{-}(q^2)$ are  transition form factors and
$P_{\mu}=(p+p')_{\mu}$, $q_{\mu}=(p-p')_{\mu}$.

From the general philosophy of the QCD sum rules, we see a hadron
from two different windows. First, we see it from the outside, so we
have a hadron with hadronic parameters such as its mass and leptonic
decay constant. Second, we see the internal structure of the hadron
namely, quarks and gluons and their interactions  in QCD vacuum. In technique language, we start with the main
object in QCD sum rules so called the correlation function. The
correlation function is calculated in two different ways: From one
side, it is saturated by a tower of hadrons called the
phenomenological or physical side. On the other hand, the QCD or
theoretical side, it is calculated in terms of quark and gluons
interacting in QCD vacuum by the help of the operator product
expansion (OPE), where the short and long distance effects are
separated. The farmer is calculated  using the perturbation theory
(perturbative contribution), however, the latter is parameterized in
terms of  vacuum condensates with different mass dimensions. In the
present work, there is  no light quarks which the non-perturbative
contributions mainly come from their vacuum condensates and  the heavy quark condensate
contributions are suppressed by inverse of the heavy quark mass and
can be safely removed. The two-gluon contributions are also very small and here, we will ignore those contributions. Hence, the only contribution comes from the perturbative part. Equating  two representations of the
correlation function and applying double Borel transformation with
respect to the momentum of the initial and final states to suppress
the contribution of the higher states and continuum, sum rules for
the physical quantities, form factors, are obtained. To proceed, we
consider the following correlation functions:
\begin{eqnarray}\label{6au}
\Pi _{\mu}(p^2,p'^2)=i^2\int d^{4}xd^4ye^{-ipx}e^{ip'y}<0\mid T[J _{
S}(y) J_{\mu}^{V;A}(0) J_{B_{c}}(x)]\mid  0>,
\end{eqnarray}
\begin{eqnarray}\label{6au1}
\Pi _{\mu\nu}(p^2,p'^2)=i^2\int d^{4}xd^4ye^{-ipx}e^{ip'y}<0\mid T[J
_{\nu AV}(y) J_{\mu}^{V;A}(0) J_{B_{c}}(x)]\mid  0>,
\end{eqnarray}
where $J _{ S}(y)=\overline{c} U c$, $J _{\nu
AV}(y)=\overline{c}\gamma_{\nu} \gamma_{5}c$,
$J_{B_{c}}(x)=\overline{b}\gamma_{5}c$ are the interpolating
currents of the S, AV and $B_c$ mesons, respectively and
$J_{\mu}^{V}(0)=~\overline{c}\gamma_{\mu}b $,
$J_{\mu}^{A}=~\overline{c}\gamma_{\mu}\gamma_{5}b$ are the the
vector and axial-vector parts of the transition current. In order to
calculate the phenomenological or physical part of the correlator given
in Eq. (\ref{6au}), two complete sets of intermediate states with
the same quantum numbers as the interpolating currents $J_{S(AV)}$
and $J_{B_{c}}$ are inserted, As a result, the following
representations of the above-mentioned correlators are obtained:
\begin{eqnarray} \label{7au}
\Pi _{\mu}(p^2,p'^2)&=&\frac{<0\mid J_{S}(0) \mid S(p')><S(p')\mid
J_{\mu}^{V;A}(0)\mid B_{c}(p)><B_{c}(p)\mid
J_{B_{c}}(0)\mid 0>}{(p'^2-m_{S}^2)(p^2-m_{B_{c}}^2)}\nonumber \\
&+& \cdots,
\end{eqnarray}
\begin{eqnarray} \label{7au1}
\Pi _{\mu\nu}(p^2,p'^2)&=&\frac{<0\mid J^{\nu}_{AV}(0) \mid
AV(p',\varepsilon)><AV(p',\varepsilon)\mid J_{\mu}^{V;A}(0)\mid
B_{c}(p)><B_{c}(p)\mid
J_{B_{c}}(0)\mid 0>}{(p'^2-m_{AV}^2)(p^2-m_{B_{c}}^2)}\nonumber \\
&+& \cdots,
\end{eqnarray}
 where $\cdots$ represents the contributions coming from higher states
 and continuum. The vacuum to the hadronic state  matrix elements in Eq. (\ref{7au}) can be
 parameterized in terms of the leptonic decay constants as:
 \begin{eqnarray}\label{8au}
<0\mid J_{S} \mid S(p')>&=&-if_{S}, ~~ <B_{c}(p)\mid J_{B_{c}}\mid
0>=-i\frac{f_{B_{c}}m_{B_{c}}^2}{m_{b}+m_{c}}, ~~ <0\mid
J^{\nu}_{AV} \mid AV(p',\varepsilon)>=f_{AV}m_{AV}\varepsilon^{\nu}.
\end{eqnarray}

 Using Eqs. (\ref{3au}-\ref{8au}), the final expressions  of the
 phenomenological side of the correlation functions are obtained as:
\begin{eqnarray}\label{9amplitude}
\Pi_{\mu}(p^2,p'^2)&=&-\frac{f_{S}}{(p'^2-m_{S}^2)(p^2-m_{B_{_{c}}}^2)}
\frac{f_{B_{c}}m_{B_{c}}^2}{m_{b}+m_{c}}\Bigg[f_{1}(q^{2}) P_{\mu}
+f_{2}(q^{2})q_{\mu}\Bigg] +\mbox{excited states,}\nonumber\\
\Pi_{\mu\nu}(p^2,p'^2)&=&\frac{f_{B_{c}}m_{B_{c}}^2}{(m_{b}+m_{c})}
\frac{f_{AV}m_{AV}} {(p'^2-m_{AV}^2)(p^2-m_{B_{c}}^2)}
\bigg[f_{0}(q^{2})g_{\mu\nu} (m_{B_{c}}+m_{AV})
-\frac{f_{+}(q^{2})P_{\mu}p_{\nu}}{(m_{B_{c}}+m_{AV})}
-\frac{f_{-}(q^{2})q_{\mu}p_{\nu}}{(m_{B_{c}}+m_{AV})}\nonumber\\
&+&\varepsilon_{\alpha\beta\mu\nu}p^{\alpha}p'^{\beta}\frac{f_{V}(q^{2})}
{(m_{B_{c}}+m_{AV})}\bigg] +\mbox{excited states,}
\end{eqnarray}
 where, we will choose the structures $P_{\mu}$,
 $q_{\mu}$, $\varepsilon_{\mu\nu\alpha\beta}p'^{\alpha}p^{\beta}$, $g_{\mu\nu}$ and $\frac{1}{2}(p_{\mu}p_{\nu}
 \pm p'_{\mu}p_{\nu})$ to evaluate the form factors $f_1$, $f_2$,
 $f_V$, $f_0$ and $f_\pm$, respectively.

On the QCD side, the aforementioned correlation functions can be
calculated by the help of the OPE in the deep space-like region
where $p^2 \ll (m_{b}+m_{c})^2 $ and $p'^2 \ll (2m_{c})^2$. As we
mentioned before, the main contributions to the theoretical part of the correlation functions come from bare-loop (perturbative) diagrams. To calculate those contributions, the correlation functions are
written in terms of the selected structures as follows:
\begin{eqnarray}
\Pi_{\mu}&=&\Pi_1^{per}P_{\mu}+\Pi^{per}_2q_{\mu},\nonumber\\
\Pi_{\mu\nu}&=&\Pi_V^{per}\varepsilon_{\mu\nu\alpha\beta}p'^{\alpha}p^{\beta}+\Pi^{per}_0g_{\mu\nu}+\frac{1}{2}\Pi^{per}_+(p_{\mu}p_{\nu}
 + p'_{\mu}p_{\nu})+\frac{1}{2}\Pi^{per}_-(p_{\mu}p_{\nu}
 - p'_{\mu}p_{\nu}),
\end{eqnarray}
where, each $\Pi_i^{per}$ function is written in terms of  the
double dispersion representation in the following way:
\begin{equation}\label{10au}
\Pi_i^{per}=-\frac{1}{(2\pi)^2}\int ds\int
ds'\frac{\rho_{i}(s,s',q^2)}{(s-p^2)(s'-p'^2)}+\textrm{ subtraction
terms},
\end{equation}
where, the functions $\rho_{i}(s,s',q^2)$ are called the spectral
densities.  Using the usual Feynman integral for the bare loop
diagram, the spectral densities  can be calculated with the help of
Cutkosky rules, i.e., by replacing the quark propagators with Dirac
delta functions: $\frac{1}{p^2-m^2}\rightarrow-2\pi\delta(p^2-m^2),$
which implies that all quarks are real. After some straightforward
calculations, the  spectral densities are obtained as follows:
\begin{eqnarray}\label{11au}
\rho_{1}(s,s',q^2)&=&N_{c}I_{0}(s,s',q^2)\Bigg[2(m_{b}-3m_{c})m_{c}+
2A\{2(m_{b}-m_{c})m_{c}-s\}
+ 2B\{2(m_{b}-m_{c})m_{c}-s^{\prime}\}\Bigg],\nonumber\\
\rho_{2}(s,s',q^2)&=&N_{c}I_{0}(s,s',q^2)\Bigg[-2(m_{b}+m_{c})m_{c}+
2A\{2(m_{b}-m_{c})m_{c}+s\} -
2B\{2(m_{b}-m_{c})m_{c}+s^{\prime}\}\Bigg],\nonumber\\
\rho_{V}(s,s',q^2)&=&4N_{c}I_{0}(s,s',q^2)\Bigg[(m_{c}-m_{b})A+2m_{c}B+m_{c}\Bigg],
\nonumber\\
\rho_{0}(s,s',q^2)&=&2N_{c}I_{0}(s,s',q^2)\Bigg[4(m_{c}^{2}-C)(m_{c}-m_{b})+m_{c}u
+\{m_{c} (4s+u)-m_{b}u\}A+2[-m_{b} s' + m_{c}(s'+u)]B \Bigg],\nonumber\\
\rho_{+}(s,s',q^2)&=&2N_{c}I_{0}(s,s',q^2)\Bigg[-m_{c}+(m_{b}-3m_{c})A-2m_{c}B
+2(m_{b}-m_{c})D+2(m_{b}-m_{c})E \Bigg]
,\nonumber \\
\rho_{-}(s,s',q^2)&=&2N_{c}I_{0}(s,s',q^2)\Bigg[m_{c}-(m_{c}+m_{b})A+2m_{c}B
+2(m_{b}-m_{c})D+2(m_{c}-m_{b})E
\Bigg],\nonumber \\
\end{eqnarray}
where
\begin{eqnarray}\label{12}
I_{0}(s,s',q^2)&=&\frac{1}{4\lambda^{1/2}(s,s',q^2)},\nonumber\\
\lambda(a,b,c)&=&a^{2}+b^{2}+c^{2}-2ac-2bc-2ab,\nonumber \\
\Delta&=&m_{b}^{2}-m_{c}^{2}-s,\nonumber \\
\Delta'&=&-s',\nonumber \\
u&=&s+s'-q^{2},\nonumber \\
A&=&\frac{1}{\lambda(s,s',q^{2})}(\Delta' u-2\Delta s'),\nonumber\\
B&=&\frac{1}{\lambda(s,s',q^{2})}(\Delta u-2\Delta' s),\nonumber\\
C&=&\frac{1}{2\lambda(s,s',q^{2})}[\Delta'^{2}s+\Delta^{2} s'-
\Delta\Delta' u+m_{c}^{2}(-4s s'+u^{2})],\nonumber\\
D&=&\frac{1}{\lambda(s,s',q^{2})^{2}}[-6 \Delta\Delta' s'
u+\Delta'^{2}(2 s s'+u^{2})
+2s'(3\Delta^{2} s'+m_{c}^{2}(-4s s'+u^{2}))],\nonumber\\
E&=&\frac{1}{\lambda(s,s',q^{2})^{2}}[-3 \Delta^{2}
s'u+2\Delta\Delta'(2 s s'+u^{2})
-u(3\Delta'^{2} s+m_{c}^{2}(-4s s'+u^{2}))],\nonumber\\
 \end{eqnarray}
and  $N_{c}=3$ is the number of colors. The integration region for
the perturbative contribution in the Eq. (\ref{10au}) is determined
requiring that the arguments of the three $\delta$ functions vanish
simultaneously. Therefore, the physical region in the $s$ and $s'$
plane is described by the following non-equality:
\begin{equation}\label{13au}
-1\leq f(s,s')=\frac{2ss'+(s+s'-q^2)(m_{b}^2-s-m_{c}^2)}
{\lambda^{1/2}(m_{b}^2,s,m_{c}^2)\lambda^{1/2}(s,s',q^2)}\leq+1.
\end{equation}

Equating the coefficient of the selected structures from the
phenomenological  and the OPE expressions and applying double Borel
transformations with respect to the variables $p^2$ and $p'^2$
($p^2\rightarrow M_{1}^2,p'^2\rightarrow M_{2}^2$) in order to
suppress the contributions of the higher states and continuum, the
QCD sum rules for the form factors $f_{1}(q^{2})$ and $f_{2}(q^{2})$
for the $B_{c}\rightarrow X_{c0}\ell\nu$ decay can be acquired:
\begin{eqnarray}\label{15au}
f_{1,2}(q^2)&=&\frac{(m_{b}+m_{c})
}{f_{B_{c}}m_{B_{c}}^2}\frac{1}{f_{X_{c0}}}e^{m_{B_{c}}^2/M_{1}^2}
e^{m_{X_{c0}}^2/M_{2}^2}\bigg\{\nonumber\\
&&\frac{1}{(2\pi)^2}\int_{(m_b+m_c)^2}^{s_0}ds
  \int_{(2m_c)^2}^{s_0'}
ds'\rho_{1,2}(s,s',q^2)\theta[1-f^{2}(s,s')]e^{-s/M_{1}^2}e^{-s'/M_{2}^2}\bigg\}.
\end{eqnarray}
 The form factors $f_{V}$, $f_{0}$, $f_{+}$ and
$f_{-}$ for $B_{c}\rightarrow AV\ell\nu$ decays are also obtained
as:
\begin{eqnarray}\label{152au}
f_{i}(q^2)&=&\kappa\frac{(m_{b}+m_{c})
}{f_{B_{c}}m_{B_{c}}^2}\frac{\eta}{f_{AV}m_{AV}}e^{m_{B_{c}}^2/M_{1}^2+
m_{AV}^2/M_{2}^2} \nonumber
\\&\times&\Bigg[\frac{1}{(2\pi)^2}\int_{4m_{c}^{2}}^{s_0'} ds' \int_{(m_{b}+m_{c})^{2}}^{s_0}
ds\rho_{i}(s,s',q^2)
\theta[1-f^{2}(s,s')]e^{-s/M_{1}^2-s'/M_{2}^2}\Bigg],
\end{eqnarray}
where $i=V,0$,$\pm$,  and $\eta=m_{B_{c}}+m_{AV}$ for $i=V,\pm $ and
$\eta=\frac{1}{m_{B_{c}}+m_{AV}}$ for $i=0$ are considered. Here
$\kappa=+1$ for $i=\pm$ and $\kappa=-1$ for $i=0$ and $V$. In the
above equations, the $s_0$ and $s'_0$ are continuum thresholds in
$s$ and $s'$ channels, respectively.

In order to subtract the contributions of the higher states and
continuum, the quark-hadron duality assumption is used, i.e., it is
assumed that
\begin{eqnarray}\label{ope}
\rho^{higher states}(s,s') = \rho^{OPE}(s,s') \theta(s-s_0)
\theta(s'-s'_0).
\end{eqnarray}

Note that, the double Borel transformation used in calculations  is
written as:
\begin{equation}\label{16au}
\hat{B}\frac{1}{(p^2-m^2_1)^m}\frac{1}{(p'^2-m^2_2)^n}\rightarrow(-1)^{m+n}\frac{1}{\Gamma(m)}\frac{1}{\Gamma
(n)}e^{-m_{1}^2/M_{1}^{2}}e^{-m_{2}^2/M_{2}^{2}}\frac{1}{(M_{1}^{2})^{m-1}(M_{2}^{2})^{n-1}}.
\end{equation}
Now, we would like to explain our reason for ignoring the contributions of the gluon condensates to the QCD side of the correlation function. These contributions for the related form factors are obtained as the following orders:
\begin{eqnarray}\label{gluon}
f_{1,2}^{\langle G^2\rangle}&\sim&\langle \frac{\alpha_s}{\pi}G^2 \rangle\frac{m_b^{n_1} m_c^{m_1}}{M_1^{2k_1} M_2^{2l_1}},~~~~~~~~~~~~~~~~~n_1+m_1=2k_1+2l_1,\nonumber\\
f_{V,+,-}^{\langle G^2\rangle}&\sim&\langle \frac{\alpha_s}{\pi}G^2 \rangle\frac{m_b^{n_2} m_c^{m_2}}{M_1^{2k_2} M_2^{2l_2}},~~~~~~~~~~~~~~~~~n_2+m_2=2k_2+2l_2-1,\nonumber\\
f_0^{\langle G^2\rangle}&\sim&\langle \frac{\alpha_s}{\pi}G^2 \rangle\frac{m_b^{n_3} m_c^{m_3}}{M_1^{2k_3} M_2^{2l_3}},~~~~~~~~~~~~~~~~~n_3+m_3=2k_3+2l_3+1,\nonumber\\
\end{eqnarray}
where, $\alpha_s$ is the strong coupling constant and $M_1^2$ and $M_2^2$ are Borel mass parameters. Recalling the magnitude of the $\langle \frac{\alpha_s}{\pi}G^2 \rangle=0.012~GeV^4$ \cite{12} and considering the working region of the Borel parameters (see numerical analysis section), the gluon condensate contributions become very small and here, we ignore those small contributions ( maximum contribution is obtained for $f_0^{\langle G^2\rangle}$, which is not more than few percent).

At the end of this section, we would like to present the
differential decay rates of the $B_{c}\rightarrow S(AV)\ell\nu$ in
terms of the transition form factors. The differential decay width
for $B_{c}\rightarrow S\ell\nu$ is obtained as follows :
\begin{eqnarray}\label{29au}
\frac{d\Gamma}{dq^2}&=&\frac{1}{192\pi^{3}m_{B_{c}}^{3}}
G_{F}^2|V_{cb}|^2\lambda^{1/2}(m_{B_{c}}^{2},m_{S}^{2},q^{2})
\left(\frac{q^{2}-m_{\ell}^{2}}{q^{2}}\right)^{2}
\left\{-\frac{1}{2}(2q^{2}+m_{\ell}^{2})\left[|f_{1}(q^{2})|^{2}
(2m_{B_{c}}^{2}+2m_{S}^{2}-q^{2})
\right.\right. \nonumber \\
&+&\left.\left.2(m_{B_{c}}^{2}-m_{S}^{2})Re[f_{1}(q^{2})f_{2}^{*}(q^{2})]+
|f_{2}(q^{2})|^{2}q^{2}\right]\right.+\left.\frac{(q^{2}+m_{\ell}^{2})}{q^{2}}\left[|f_{1}(q^{2})|^{2}(m_{B_{c}}^{2}-
m_{S}^{2})^{2}
\right.\right. \nonumber \\
&+&\left.\left.2(m_{B_{c}}^{2}-m_{S}^{2})q^{2}Re[f_{1}(q^{2})f_{2}^{*}(q^{2})]+
|f_{2}(q^{2})|^{2}q^{4}\right] \right\},
\end{eqnarray}
and also, the differential decay width corresponding to
$B_{c}\rightarrow AV\ell\nu$ decays are acquired as:
\begin{eqnarray}\label{281au}
\frac{d\Gamma}{dq^2}&=&\frac{1}{16\pi^4m_{B_{c}}^2}|\overrightarrow{p'}|
G_{F}^2|V_{cb}|^2\Bigg(4\Bigg\{(2A_{1}+A_{2}q^2)[\mid
f_{V}\mid^2(4m_{B_{c}}^2\mid\overrightarrow{p'}\mid^2)+\mid
f_{0}\mid^2]\Bigg\}\nonumber\\
&+&\left\{(2A_{1}+A_{2}q^2)\Bigg[\mid
f_{V}\mid^2(4m_{B_{c}}^2\mid\overrightarrow{p'}\mid^2 \right.
 +
m_{B_{c}}^2\frac{\mid\overrightarrow{p'}\mid^2}{m_{AV}^{2}}
(m_{B_{c}}^2-m_{AV}^2-q^2))\nonumber \\
&+&\mid f_{0}\mid^2 - \mid
f_{+}\mid^2\frac{m_{B_{c}}^2\mid\overrightarrow{p'}\mid^2}{m_{AV}^2}
(2m_{B_{c}}^2 +2m_{AV}^2 -q^2)-\mid
f_{-}\mid^2\frac{m_{B_{c}}^2\mid\overrightarrow{p'}\mid^2}{m_{AV}^2}q^2
\nonumber\\&-& 2 \left.
\frac{m_{B_{c}}^2\mid\overrightarrow{p'}\mid^2}{m_{AV}^2}(Re(f'_{0}
f_{+}+f'_{0} f_{-}+(m_{B_{c}}^2-m_{AV}^2)f_{+}f_{-}))\right]
\nonumber
\\ &-&
2A_{2}\frac{m_{B_{c}}^2\mid\overrightarrow{p'}\mid^2}{m_{AV}^2}
\Bigg[\mid f_{0}\mid^2+(m_{B_{c}}^2-m_{AV}^2)^2\mid
f_{+}\mid^2+q^4\mid f_{-}\mid^2  \nonumber \\ &+& 2(
\left.m_{B_{c}}^2-m_{AV}^2)Re(f_{0}f_{+})
+2q^2f_{0}f_{-}+2q^2(m_{B_{c}}^2-m_{AV}^2)Re(f_{+}f_{-})
\Bigg]\right\}\Bigg),
\nonumber \\
\end{eqnarray}
where\\
\begin{eqnarray}\label{30au}
\mid\overrightarrow{p'}\mid&=&\frac{\lambda^{1/2}(m_{B_{c}}^2,m_{AV}^2,q^2)}
{2m_{B_{c}}},
\nonumber\\
A_{1}&=&\frac{1}{12q^2}(q^2-m_{l}^2)^2I'_{0},\nonumber\\
A_{2}&=&\frac{1}{6q^4}(q^2-m_{l}^2)(q^2+2m_{l}^2)I'_{0},\nonumber\\
I'_{0}&=&\frac{\pi}{2}(1-\frac{m_{l}^2}{q^2}).\nonumber\\
\end{eqnarray}
%
\section{Heavy Quark Effective Theory Limit of the Form Factors }

In this section, we calculate the  heavy quark effective theory (HQET) limits of the transition form factors for
$B_c\rightarrow S(AV)\ell\nu$. For this aim, following  references \cite{ming,neubert1,kazem,kazemmel}, we use the 
parameterization
\begin{equation}\label{melau}
 y=\nu\nu'=\frac{m_{B_{c}}^2+m_{S(AV)}^2-q^2}{2m_{B_{c}}m_{S(AV)}},
 \end{equation}
 where $\nu$ and $\nu'$ are
  the four-velocities of the initial and final meson states, respectively. Next, we try to find the y dependent
  expressions of the form factors by taking
  $m_{b}\rightarrow\infty$, $m_{c}=\frac{m_{b}}{\sqrt{z}}$, where
  z is given by  $\sqrt{z}=y+\sqrt{y^2-1}$.
  In this limit,  the new Borel
  parameters  $T_{1}=M_{1}^{2}/2  (m_{b}+m_{c})$ and $T_{2}=M_{2}^{2}/4 
  m_{c}$ are defined. The new continuum thresholds $\nu_{0}$, and $\nu_{0}'$ are also parameterized  as:
\begin{equation}\label{17au}
 \nu_{0}=\frac{s_{0}-(m_{b}+m_{c})^2}{m_{b}+m_{c}},~~~~~~
 \nu'_{0}=\frac{s'_{0}-4m_{c}^2}{2m_{c}},
 \end{equation}
 and the new integration variables take the following form:
 \begin{equation}\label{18au}
 \nu=\frac{s-(m_{b}+m_{c})^2}{m_{b}+m_{c}},~~~~~~ \nu'=\frac{s'-4 m_{c}^2}{2m_{c}}.
 \end{equation}
 The leptonic decay constants are rescaled:
 \begin{equation}\label{21au}
\hat{f}_{B_{c}}=\sqrt{m_{b}+m_{c}}
f_{B_{c}},~~~~~~~\hat{f}_{S(AV)}=\sqrt{2 m_{c}} f_{S(AV)}.
\end{equation}
To evaluate the form factors in HQET, we also need to redefine the form factors in the following form:
\begin{eqnarray}\label{redefinitionha}
f'_{1,2}&=&\frac{f_{1,2}}{(m_{B_c}+m_{S})^2}\nonumber\\
f'_{V,0,+,-}&=&\frac{f_{V,0,+,-}}{m_{B_c}+m_{AV}}.
\end{eqnarray}

After  standard calculations, we obtain the y-dependent
expressions of the form factors for $B_{c}\rightarrow S\ell\nu$
transition as follows:
\begin{eqnarray}\label{22au1}
f'_{1}&=&\frac{3(-1+y^2)\Bigg[3+z+y(-3-2\sqrt{z}+z)\Bigg]}{8\sqrt{2}\pi^2
\hat{f}_{S}\hat{f}_{B_{c}}z^{13/4}(1+y)\sqrt{1+\frac{1}{\sqrt{z}}}\Bigg[\frac{(-1+y^2)
(1+\sqrt{z})^{2}}{z^2}\Bigg]^{\frac{3}{2}}}e^{(\frac{\Lambda}{T_{1}}
+\frac{\overline{\Lambda}}{T_{2}})}\Bigg\{
\int_{0}^{\nu_{0}}d\nu\int_{0}^{\nu_{0}'}d\nu'
e^{-\frac{\nu}{2T_{1}}-\frac{\nu'}{2T_{2}}}\nonumber\\&\times&\theta[1-lim_{m_{b}\rightarrow\infty}f^{2}(v,v')]\Bigg\}
,\nonumber\\
\end{eqnarray}
\begin{eqnarray}\label{222au2}
f'_{2}&=&\frac{-3(-1+y^2)\Bigg[-1+y(1+\sqrt{z})^2+4\sqrt{z}+z)\Bigg]}{8\sqrt{2}\pi^2
\hat{f}_{S}\hat{f}_{B_{c}}z^{13/4}(1+y)\sqrt{1+\frac{1}{\sqrt{z}}}
\Bigg[\frac{(-1+y^2)
(1+\sqrt{z})^{2}}{z^2}\Bigg]^{\frac{3}{2}}}e^{(\frac{\Lambda}{T_{1}}
+\frac{\overline{\Lambda}}{T_{2}})}\Bigg\{
\int_{0}^{\nu_{0}}d\nu\int_{0}^{\nu_{0}'}d\nu'
e^{-\frac{\nu}{2T_{1}}-\frac{\nu'}{2T_{2}}}\nonumber\\&\times&\theta[1-lim_{m_{b}\rightarrow\infty}f^{2}(v,v')]\Bigg\}
,\nonumber\\
\end{eqnarray}
and for $B_{c}\rightarrow AV\ell\nu$ decay, the y-dependent
expressions of the form factors are acquired  as:
\begin{eqnarray}\label{22au3}
f'_{V}&=&\frac{3(3+\sqrt{z})\Bigg[-1+y+\sqrt{z}+y\sqrt{z}\Bigg]}{8\sqrt{2}\pi^2
\hat{f}_{AV}\hat{f}_{B_{c}}z^{5/4}(1+y)(1+\sqrt{z})\sqrt{1+\frac{1}{\sqrt{z}}}
\sqrt{\frac{(-1+y^2)(1+\sqrt{z})^2}{z^2}}}e^{(\frac{\Lambda}{T_{1}}
+\frac{\overline{\Lambda}}{T_{2}})}\Bigg\{
\int_{0}^{\nu_{0}}d\nu\int_{0}^{\nu_{0}'}d\nu'
e^{-\frac{\nu}{2T_{1}}-\frac{\nu'}{2T_{2}}}\nonumber\\&\times&\theta[1-lim_{m_{b}\rightarrow\infty}f^{2}(v,v')]\Bigg\}
,\nonumber\\
\end{eqnarray}
\begin{eqnarray}\label{22au}
f'_{0}&=&\frac{3(-1+y)\Bigg[1+2y(1+\sqrt{z})+3\sqrt{z}\Bigg]}{8\sqrt{2}\pi^2
\hat{f}_{AV}\hat{f}_{B_{c}}z^{5/4}(1+y)(3+\sqrt{z})\sqrt{1+\frac{1}{\sqrt{z}}}
\sqrt{\frac{(-1+y^2)(1+\sqrt{z})^2}{z^2}}}e^{(\frac{\Lambda}{T_{1}}
+\frac{\overline{\Lambda}}{T_{2}})}\Bigg\{
\int_{0}^{\nu_{0}}d\nu\int_{0}^{\nu_{0}'}d\nu'
e^{-\frac{\nu}{2T_{1}}-\frac{\nu'}{2T_{2}}}\nonumber\\&\times&\theta[1-lim_{m_{b}\rightarrow\infty}f^{2}(v,v')]\Bigg\}
,\nonumber\\
\end{eqnarray}
\begin{eqnarray}\label{22au}
f'_{+}&=&\frac{3(-1+y^2)(3+\sqrt{z})\Bigg[2+2y^2(1+\sqrt{z})^2+5y(-1+z)-10\sqrt{z}\Bigg]}
{32\sqrt{2}\pi^2
\hat{f}_{AV}\hat{f}_{B_{c}}z^{13/4}(1+y)^2\sqrt{1+\frac{1}{\sqrt{z}}}
\Bigg[\frac{(-1+y^2)(1+\sqrt{z})^2}{z^2}\Bigg]^{\frac{3}{2}}}e^{(\frac{\Lambda}{T_{1}}
+\frac{\overline{\Lambda}}{T_{2}})}\Bigg\{
\int_{0}^{\nu_{0}}d\nu\int_{0}^{\nu_{0}'}d\nu'
e^{-\frac{\nu}{2T_{1}}-\frac{\nu'}{2T_{2}}}\nonumber\\&\times&\theta[1-lim_{m_{b}\rightarrow\infty}f^{2}(v,v')]\Bigg\}
,\nonumber\\
\end{eqnarray}
\begin{eqnarray}\label{22au}
f'_{-}&=&\frac{-3(-1+y^2)(3+\sqrt{z})\Bigg[-2+2y^2(1+\sqrt{z})^2+y(3+8\sqrt{z}+5z)
+10\sqrt{z}\Bigg]} {32\sqrt{2}\pi^2
\hat{f}_{AV}\hat{f}_{B_{c}}z^{13/4}(1+y)^2\sqrt{1+\frac{1}{\sqrt{z}}}
\Bigg[\frac{(-1+y^2)(1+\sqrt{z})^2}{z^2}\Bigg]^{\frac{3}{2}}}e^{(\frac{\Lambda}{T_{1}}
+\frac{\overline{\Lambda}}{T_{2}})}\Bigg\{
\int_{0}^{\nu_{0}}d\nu\int_{0}^{\nu_{0}'}d\nu'
e^{-\frac{\nu}{2T_{1}}-\frac{\nu'}{2T_{2}}}\nonumber\\&\times&\theta[1-lim_{m_{b}\rightarrow\infty}f^{2}(v,v')]\Bigg\}
,\nonumber\\
\end{eqnarray}
where $\Lambda=m_{B_{c}}-(m_{b}+m_{c})$ and
$\bar{\Lambda}=m_{S(AV)}-2m_{c}$.

%

\section{Numerical analysis}

This section is devoted to the numerical analysis of the form
factors, their HQET limit and branching ratios. The sum rules expressions for the form
factors depict that they mainly depend on the leptonic decay
constants, continuum thresholds $s_{0}$ and $s'_{0} $ and Borel
parameters $M_{1}^2$ and $M_{2}^2$. In calculations, the quark
masses are taken to be $ m_{c}(\mu=m_{c})=
 1.275\pm0.015~ GeV$, $m_{b} =(4.7\pm0.1)~GeV$ \cite{20} and
 the meson masses are chosen as: $m_{B_{c}}=6.286~GeV$,
 $m_{h_{c}}=3.52528~GeV$, $m_{X_{c0}}=3.41476~GeV$,
 $m_{X_{c1}}=3.51066~GeV$ \cite{16}. For
 the values of the leptonic decay
constants, we use $f_{B_{c}} = (400\pm
 40~) ~MeV $ and $f_{X_{c0}}= f_{X_{c1}}=f_{h_{c}}=(340^{+119}_{-101})
  ~MeV $\cite{Wang}.  The
two-point QCD sum rules are used to determine the continuum
thresholds $s_{0}$ and $s_{0}' $. These thresholds are not
completely arbitrary and they are related to the energy of the
exited states. The result of the physical quantities, form factors,
should be stable with respect to the small variation of these
parameters. Generally, the $s_0$ are obtained to be
$(m_{hadron}+0.5)^2$ \cite{12}. Here, we use $s_{0} =(45\pm5)~
GeV^2$ and $s_{0}' =(16\pm2)~ GeV^2 $. Since the Borel parameters
$M_{1}^2$ and $M_{2}^2 $ are not physical quantities, the form
factors should not depend on them. The reliable regions for the
Borel parameters $M_{1}^2 $ and $M_{2}^2$ can be determined by
requiring that not only the contributions of the higher states and
continuum are effectively suppressed, but  the contribution of the
operator with the highest dimension be small. As a result of the
above-mentioned requirements, the working regions are determined to
be $ 15~ GeV^2 \leq M_{1}^2 \leq35~ GeV^2 $ and $ 10~ GeV^2\leq
M_{2}^2 \leq20 ~GeV^2$. The numerical values of the form factors at
$q^2=0$  for $B_{c}\rightarrow X_{c0}\ell\nu$ and  $B_{c}\rightarrow
AV\ell\nu$ transitions are  given in the Tables \ref{tab:f1} and
\ref{tab:f2}, respectively.
\begin{table}[h]
\centering
\begin{tabular}{|c|c|c|} \hline
& $f_{1}(0)$  & $f_{2}(0)$  \\\cline{1-3}
 $B_{c}\rightarrow X_{c0}\ell\nu$ & $0.673\pm0.195$ & $-1.458 \pm 0.437$ \\\cline{1-3}

 \end{tabular}
 \vspace{0.8cm}
\caption{The values of the form factors for the $B_{c}\rightarrow
X_{c0}\ell\nu$} decay at $M_{1}^2=25~GeV^2$, $M_{2}^2=15~GeV^2$ and
$q^{2}=0$. \label{tab:f1}
\end{table}

\begin{table}[h]
\centering
\begin{tabular}{|c|c|c|c|c|} \hline
& $f_{0}(0)$  & $f_{V}(0)$ &  $f_{+}(0)$ &  $f_{-}(0)$
\\\cline{1-5}
 $B_{c}\rightarrow X_{c1}\ell\nu$ & $0.084\pm 0.025$& $0.949\pm0.261$& $0.211\pm0.061 $ &$ -0.586\pm0.179 $\\\cline{1-5}
 $B_{c}\rightarrow h_{c}\ell\nu$ & $0.084\pm 0.025 $& $0.954\pm 0.282$& $0.211\pm0.061 $ &$ -0.588\pm0.181$\\\cline{1-5}
 \end{tabular}
 \vspace{0.8cm}
\caption{The values of the form factors for the $B_{c}\rightarrow
AV\ell\nu$} decays at $M_{1}^2=25~GeV^2$, $M_{2}^2=15~GeV^2$ and
$q^{2}=0$. \label{tab:f2}
\end{table}
\begin{table}[h]
\centering
\begin{tabular}{|c|c|c|c|} \hline
& a  & b & $m_{fit}$\\\cline{1-4}
 $f_{1}(B_{c}\rightarrow X_{c0}\ell\nu)$ & 0.218 & 0.455 & 5.043\\\cline{1-4}
 $f_{2}(B_{c}\rightarrow X_{c0}\ell\nu)$ & -0.721  & -0.738 & 4.492\\\cline{1-4}
 \end{tabular}
 \vspace{0.8cm}
\caption{Parameters appearing in the form factors of the
$B_{c}\rightarrow X_{c0}\ell\nu$} decay at $M_{1}^2=25~GeV^2$ and
$M_{2}^2=15~GeV^2$. \label{tab:1}
\end{table}

\begin{table}[h]
\centering
\begin{tabular}{|c|c|c|c|} \hline
& a  & b & $m_{fit}$\\\cline{1-4}
 $f_{0}(B_{c}\rightarrow X_{c1}\ell\nu)$ & 0.211 & -0.126 & 5.241\\\cline{1-4}
 $f_{V}(B_{c}\rightarrow X_{c1}\ell\nu)$ & 0.512  & 0.438 & 4.711\\\cline{1-4}
 $f_{+}(B_{c}\rightarrow X_{c1}\ell\nu)$ & 0.279  & -0.068 & 3.872\\\cline{1-4}
 $f_{-}(B_{c}\rightarrow X_{c1}\ell\nu)$ & -0.594  & 0.008 & 3.735\\\cline{1-4}
 $f_{0}(B_{c}\rightarrow h_{c}\ell\nu)$ & 0.211  & -0.127 & 5.256\\\cline{1-4}
 $f_{V}(B_{c}\rightarrow h_{c}\ell\nu)$ & 0.498  & 0.456 & 4.735\\\cline{1-4}
 $f_{+}(B_{c}\rightarrow h_{c}\ell\nu)$ & 0.282  & -0.702 & 3.839\\\cline{1-4}
 $f_{-}(B_{c}\rightarrow h_{c}\ell\nu)$ & -0.620  & 0.031 &
3.686\\\cline{1-4}
 \end{tabular}
 \vspace{0.8cm}
\caption{Parameters appearing in the form factors of the
$B_{c}\rightarrow X_{c1}\ell\nu$} and $B_{c}\rightarrow
h_{c}\ell\nu$ decays at $M_{1}^2=25~GeV^2$ and $M_{2}^2=15~GeV^2$.
\label{tab:2}
\end{table}
\begin{table}[h]
\centering
\begin{tabular}{|c||c|c|c|c|c|c|c|c|} \hline
 $q^2 (GeV^{2})$ & 0 & 1 & 2& 3& 4&5&6&7\\\cline{1-9}
 y& 1.1920 & 1.1687 & 1.1454& 1.1221& 1.0988&1.0755&1.0522&1.0289\\\cline{1-9}
  \hline \hline
 $f_{1}$ & 0.6735 & 0.7204 & 0.7732& 0.8326& 0.9001&0.9771&1.0656&1.1680\\\cline{1-9}
 $f_{1}(HQET)$ & 0.3423 & 0.3614 & 0.4087& 0.4637& 0.5483&0.6523&0.7833&0.9432\\\cline{1-9}
 $f_{2}$ & -1.4594  & -1.5760 & -1.7102& -1.8658& -2.0480&-2.2636&-2.5218&-2.8354\\\cline{1-9}
 $f_{2}(HQET)$ & -0.8921   & -0.9432&-1.0824& -1.1841& -1.3682&-1.6112&-2.0571&-2.4633\\\cline{1-9}

 \end{tabular}
 \vspace{0.8cm}
\caption{Values of the   form factors and their HQET limits for the
$B_{c}\rightarrow X_{c0}\ell\nu$ at $M_{1}^{2}=25 ~GeV^{2}$,
$M_{2}^{2}=15~ GeV^{2}$, $T_{1}=2.09~ GeV$ and $T_{2}=2.94~
GeV$.} \label{tab:41}
\end{table}
\begin{table}[h]
\centering
\begin{tabular}{|c||c|c|c|c|c|c|c|c|} \hline
 $q^2 (GeV^{2})$ & 0 & 1 & 2& 3& 4&5&6&7\\\cline{1-9}
 y& 1.1745 & 1.1518 &1.1292& 1.1065& 1.0838&1.0612&1.0385&1.0159\\\cline{1-9}
  \hline \hline
 $f_{0}$ & 0.0841 & 0.0823 & 0.0800& 0.0770& 0.0732&0.0684&0.0624&0.0548\\\cline{1-9}
 $f_{0}(HQET)$ & 0.0683 & 0.0675 & 0.0664& 0.0652& 0.0641&0.0628&0.0604&0.0559\\\cline{1-9}
 $f_{V}$ & 0.9506 & 1.0171& 1.0925& 1.1784& 1.2771&1.3915&1.5252&1.6833\\\cline{1-9}
 $f_{V}(HQET)$ & 0.4739  & 0.5421&0.6331& 0.7566& 0.9054&1.0872&1.1543&1.3421\\\cline{1-9}
 $f_{-}$ &-0.5862 & -0.6309 & -0.6828& -0.7442& -0.8176&-0.9071&-1.0185&-1.1612\\\cline{1-9}
 $f_{-}(HQET)$ &-0.2954 & -0.3264 & -0.3682& -0.4448& -0.5412&-0.6518&-0.7839&-1.0173\\\cline{1-9}
 $f_{+}$ & 0.2108 & 0.2207 & 0.2312& 0.2424& 0.2539&0.2654&0.2761&0.2841\\\cline{1-9}
 $f_{+}(HQET)$ & 0.1032 & 0.1157&0.1302& 0.1545& 0.1771&0.1998&0.2152&0.2305\\\cline{1-9}

 \end{tabular}
 \vspace{0.8cm}
\caption{Values of the   form factors and their HQET limits for the
$B_{c}\rightarrow X_{c1}\ell\nu$ at $M_{1}^{2}=25~GeV^{2}$,
$M_{2}^{2}=15 ~GeV^{2}$, $T_{1}=2.09 ~GeV$ and $T_{2}=2.94 ~GeV$.}
\label{tab:42}
\end{table}
\begin{table}[h]
\centering
\begin{tabular}{|c||c|c|c|c|c|c|c|c|} \hline
 $q^2 (GeV^{2})$ & 0 & 1 & 2& 3& 4&5&6&7\\\cline{1-9}
 y& 1.1745 & 1.1518 &1.1292& 1.1065& 1.0838&1.0612&1.0385&1.0159\\\cline{1-9}
  \hline \hline
 $f_{0}$ & 0.0842& 0.0824& 0.0801& 0.0771& 0.0733&0.0685&0.0625&0.0550\\\cline{1-9}
 $f_{0}(HQET)$ & 0.0692 & 0.0683 & 0.0665& 0.0653& 0.0641&0.0629&0.0604&0.0561\\\cline{1-9}
 $f_{V}$ & 0.9545& 1.0213 & 1.0970&1.1833& 1.2824&1.3970&1.5310&1.6890\\\cline{1-9}
 $f_{V}(HQET)$ &0.4781& 0.5483 & 0.6383& 0.7627& 0.9061&1.0922&1.1633&1.3948\\\cline{1-9}
 $f_{-}$ &-0.5891& -0.6332 &-0.6845& -0.7448& -0.8167&-0.9038&-1.0114&-1.1477\\\cline{1-9}
 $f_{-}(HQET)$ &-0.2983& -0.3291& -0.3704& -0.4457& -0.5404&-0.6487&-0.7671&-1.0102\\\cline{1-9}
 $f_{+}$ & 0.2117 &0.2217 &0.2322& 0.2433& 0.2547&0.2659&0.2758&0.2823\\\cline{1-9}
 $f_{+}(HQET)$ & 0.1043& 0.1166&0.1314&0.1557& 0.1783& 0.2017&0.2163&0.2314\\\cline{1-9}

 \end{tabular}
 \vspace{0.8cm}
\caption{Values of the  form factors and their HQET limits for the
$B_{c}\rightarrow h_{c}\ell\nu$ at $M_{1}^{2}=25 ~GeV^{2}$,
$M_{2}^{2}=15~ GeV^{2}$, $T_{1}=2.09~ GeV$ and $T_{2}=2.94~GeV$.}
\label{tab:43}
\end{table}

In order to estimate the decay width of the $B_{c}\rightarrow S(AV)
\ell\nu$ transitions, we need to know the $q^2$ dependent  form
factors  in the whole physical region, $ m_{l}^2 \leq q^2 \leq
(m_{B_{c}} - m_{S(AV)})^2$. Our form factors are truncated at about
$q^2 =4 ~GeV^2$. To extend our results to the full physical region,
we search for parameterization of the form factors in such a way that
in the region $0 \leq q^2 \leq 4~ GeV^2$, this parameterization
coincides with the sum rules predictions. The following fit
parameterization is chosen  for the form factors with respect to
$q^2$:
\begin{equation}\label{172au}
f_{i}(q^2)=\frac{a}{(1- \frac{q^{2}}{m_{fit}^{2}})}+\frac{b}{(1-
\frac{q^{2}}{m_{fit}^{2}})^{2}},
\end{equation}
where, the values of the parameters $a$, $b$ and $m_{fit}$ for the
$B_{c}\rightarrow X_{c0}\ell\nu$ and $B_{c}\rightarrow
(X_{c1},h_{c})\ell\nu$  are given in the Tables \ref{tab:1} and
\ref{tab:2}, respectively.
To calculate the numerical values of the form factors at
HQET limit, the values of 
 $\Lambda=0.31 GeV$ and 
$\overline{\Lambda}=0.86GeV (0.96GeV)$ are used for
$B_{c}\rightarrow S\ell\nu$ ($B_{c}\rightarrow
AV\ell\nu$) transitions,
respectively (see \cite{huang,dai}). In Tables,  \ref{tab:41},  \ref{tab:42} and \ref{tab:43}, we compare the values of the form factors and their HQET limits for considered transitions in the interval $0\leq q^2\leq7$ and corresponding values of the y.  Comparing the form factors and their HQET values in those Tables, we see that all form factors and their HQET limits have the same behavior with respect to the $q^2$, i.e., they both growth or fail by increasing the values of $q^2$.  The HQET limit of the form factors  are comparable with their original values and in large $q^2$, those form factors and their HQET values become very close to  each other. The results presented at   Tables,  \ref{tab:42} and \ref{tab:43} also indicate that  the  form factors and their HQET limits  for $B_{c}\rightarrow
X_{c1}\ell\nu$ and  $B_{c}\rightarrow
h_{c}\ell\nu$ have values very close to each other since the $X_{c1}$ and $h_{c}$ mesons are both axial vectors, i.e., $J^P=1^+$ and have nearly the same mass.

At the end of this section, we would like to calculate the values of
the branching ratios for these decays. Taking into account the $q^2$
dependency of the form factors and performing integration over $q^2$
from the differential decay rates in Eqs. (\ref{29au}, \ref{281au})
in the interval $m_{l}^2\leq q^2\leq(m_{B_{c}}-m_{S(AV)})^2$ and
also using the total life-time of the $B_c$ meson
$\tau_{B_{c}}=0.46\pm0.07\times10^{-12}s$ \cite{16}, we obtain the
branching ratios of the related transitions  as presented in Table
\ref{tab:3}.
\begin{table}[h]
\centering
\begin{tabular}{|c|c|c|c|} \hline
&  $B_{c}\rightarrow X_{c0}\ell\nu$ & $B_{c}\rightarrow
X_{c1}\ell\nu$ & $B_{c}\rightarrow h_{c}\ell\nu$\\\cline{1-4}
 Present work & 0.182$\pm$0.051 & 0.146$\pm$0.042 & 0.142$\pm$0.040\\\cline{1-4}
 CLQM \cite{Wang} & $0.21^{+0.02+0.01}_{-0.04-0.01}$  &
$0.14^{+0.00+0.01}_{-0.01-0.01}$ &
$0.31^{+0.05+0.01}_{-0.08-0.01}$\\\cline{1-4}
 RGM~\cite{Chang:2001pm}  & 0.12  & 0.15 & 0.18\\\cline{1-4}
 RCQM~\cite{Ivanov:2006ni} & 0.17  & 0.092 & 0.27\\\cline{1-4}
 RCQM~\cite{Ivanov:2005fd} & 0.18  & 0.098 & 0.31\\\cline{1-4}
 NRCQM\cite{Hernandez:2006gt} & 0.11  & 0.066 & 0.17\\\cline{1-4}
 \hline \hline
&  $B_{c}\rightarrow X_{c0}\tau\nu$ & $B_{c}\rightarrow
X_{c1}\tau\nu$ & $B_{c}\rightarrow h_{c}\tau\nu$\\\cline{1-4}
 Present work & 0.049$\pm$0.016 & 0.0147$\pm$0.0044 & 0.0137$\pm$0.0038\\\cline{1-4}
 CLQM \cite{Wang} & $0.024^{+0.001+0.001}_{-0.003-0.001}$  &
$0.015^{+0.000+0.001}_{-0.001-0.002}$ &
$0.022^{+0.002+0.000}_{-0.004-0.000}$\\\cline{1-4}
 RGM~\cite{Chang:2001pm}  & 0.017  & 0.024 & 0.025\\\cline{1-4}
 RCQM~\cite{Ivanov:2006ni} & 0.013  & 0.0089 & 0.017\\\cline{1-4}
 RCQM~\cite{Ivanov:2005fd} & 0.018  & 0.012 & 0.027\\\cline{1-4}
 NRCQM\cite{Hernandez:2006gt} & 0.013  & 0.0072 & 0.015\\\cline{1-4}
  \end{tabular}
 \vspace{0.8cm}
\caption{Branching ratios of the semileptonic  $B_{c}\rightarrow
(X_{c0}, X_{c1}, h_{c})\ell\nu$ $(\ell=e,\mu,\tau)$ transitions in
different approaches. } \label{tab:3}
\end{table}
This Table is also contain the  predictions  of the other approaches
such as covariant light-front quark model (CLQM), renormalization
group method (RGM), relativistic constituent quark model (RCQM) and
nonrelativistic constituent quark model (NRCQM) \cite{Wang,
Chang:2001pm, Ivanov:2006ni, Ivanov:2005fd, Hernandez:2006gt}. These
results can be tested in the future experiments.

In conclusion, using the QCD sum rules approach, we investigated the
semileptonic $B_c \rightarrow S(AV) \ell \nu$ decays. The $q^2$
dependencies of the transition form factors were calculated. The HQET limits of the form factors were also evaluated and compared with original form factors. The
obtained results were used to estimate the total decay widths and
branching ratios of these transitions.  A comparison of the results
for branching fractions was also presented.

\section{Acknowledgment}
  The authors would like to thank T. M. Aliev and A. Ozpineci for
  their useful discussions. K. A.  would like to thank TUBITAK, Turkish Scientific and Research
Council, for their partial financial support.


\clearpage
\begin{thebibliography}{99}
%
\bibitem{1} F. Abe et al., CDF Collabration, {\it Phys. Rev.} {\bf D 58} 112004
(1998).
%
\bibitem{2} A. Adb El-Hady, M. A. K. Lodhi and J. P. Vary, {\it Phys. Rev.} {\bf D 59}
094001 (1999).
%
\bibitem{3} L P Fulcher, {\it Phys. Rev.} {\bf D 60}
074006 (1999).
%
\bibitem{4} D. Ebert, R. N. Faustov and V. O. Galkin, {\it Phys. Rev.} {\bf D 67}
014027 (2003).
%
\bibitem{5} S. Godfrey, {\it Phys. Rev.} {\bf D 70}
054017 (2004).
%
\bibitem{6} A. K. Rai, P. C.  Vinodkumar, {\it Pramana 66}, 953 (2006), {\bf hep-ph/0606194v1}.
%
\bibitem{7} S. Stone, to appear in proceedings of
 "Heavy Flavor Physics: A Probe of Nature's Grand Design",
 Varenna, Italy, July 1997, {\bf hep-ph/9709500}.
%
\bibitem{8} D.S. Du, Z. Wang, {\it Phys. Rev.} {\bf D 39}
1342 (1989); C.H. Chang, Y.Q. Chen, ibid. {\bf 48} (1993) 4086; K.
Cheung, {\it Phys. Rev. Lett.} {\bf 71} 3413 (1993); E. Braaten, K.
Cheung, T. Yuan, {\it Phys. Rev.} {\bf D 48} R5049 (1993)
%
\bibitem{Aliev1}   T. M. Aliev, M. Savci,  Phys. Lett. {\bf B 434} (1998) 358.
%
\bibitem{Aliev2} T. M. Aliev, M. Savci,  J. Phys. {\bf G 24} (1998) 2223.
%
\bibitem{Aliev3}   T. M. Aliev, M. Savci, Eur. Phys. J. {\bf C 47} (2006)
413.
%
\bibitem{Alievsp}   T. M. Aliev, M. Savci,  Phys. Lett. {\bf B 480} (2000)
97.
%
\bibitem{Kazým1} N. Ghahramany, R. Khosravi, K. Azizi {\it Phys. Rev.} {\bf D 78}
116009 (2008).
%
\bibitem{Kazým2}  K. Azizi, R. Khosravi {\it Phys. Rev.} {\bf D 78}
036005 (2008).
%
\bibitem{Kazým3} K. Azizi, F. Falahati, V. Bashiry, S. M. Zebarjad {\it Phys. Rev.} {\bf D 77}
114024 (2008).
%
\bibitem{Kazým4} K. Azizi,  R. Khosravi, V. Bashiry,  {\it Eur. Phys. J.} {\bf C 56}
357-370 (2008).
%
\bibitem{Kazým5} K. Azizi, V. Bashiry,  {\it Phys. Rev.} {\bf D 76}
114007 (2007).
%
\bibitem{yuming} Yu-Ming Wang, Cai-Dian Lu  {\it Phys. Rev.} {\bf D 77}, 054003 (2008).
\bibitem{Ivanov} M. A. Ivanov, J. G. Korner and P. Santorelli,  {\it Phys. Rev.} {\bf D 73}
054024 (2006).
\bibitem{galkin1} D. Ebert, R.N. Faustov, V.O. Galkin,  {\it Phys. Rev.} {\bf D 68},
094020 (2003).
\bibitem{galkin2} D. Ebert, R.N. Faustov, V.O. Galkin,  {\it  Eur. Phys. J.} {\bf C 32},
29 (2003).
\bibitem{Faessler} A. Faessler, Th. Gutsche, M. A. Ivanov, J. G. K¨orner,
V. E. Lyubovitskij,  {\it Eur. Phys. J.direct } {\bf C 4}, 18(2002).

\bibitem{Wang} Xiao-Xia Wang, Wei Wang, Cai-Dian Lü, {\bf hep-ph/0901.193v1}.
\bibitem{Chang:2001pm} C.~H.~Chang, Y.~Q.~Chen, G.~L.~Wang and H.~S.~Zong, {\it Phys. Rev.} {\bf D 65}, 014017
(2002).
  %

\bibitem{Ivanov:2006ni} M.~A.~Ivanov, J.~G.~Korner and P.~Santorelli, {\it Phys. Rev.}  {\bf D 73}, 054024
(2006).
  %
\bibitem{Ivanov:2005fd}
  M.~A.~Ivanov, J.~G.~Korner and P.~Santorelli,  {\it Phys. Rev.}   {\bf D 71}, 094006 (2005)
  [Erratum-ibid.\   {\bf D 75}, 019901 (2007)].
%
\bibitem{Hernandez:2006gt}
  E.~Hernandez, J.~Nieves and J.~M.~Verde-Velasco,
    {\it Phys. Rev.}   {\bf D 74}, 074008 (2006)
  [arXiv:hep-ph/0607150].

%
\bibitem{BcXchc1} T. Becher, H. Boos, E. Lunghi, {\it JHEP0712:062} (2007).
%
\bibitem{BcXchc2} U. Aglietti, L. D. Giustino, G. Ferrera, A. Renzaglia,
G. Ricciardi, L. Trentadue,  {\it Phys. Lett.} {\bf B 653} 38-52
(2007).
%
\bibitem{BcXchc3} P. Urquijo, Belle Collaboration, {\it Proceedings of the
XXXIII International Conference of the High Energy Physics
(ICHEP'06), Moscow}, (2006), arXiv: hep-ex/0611049.

%
\bibitem{BcXchc4} K. Abe, Belle Collaboration, {\it Proceedings of the
XXXIII International Conference of the High Energy Physics
(ICHEP'06), Moscow}, (2006), arXiv: hep-ex/0609013.
%
\bibitem{BcXchc5} C. Schwanda, Belle Collaboration,  {\it Phys. Rev.} {\bf D 75}
032005 (2007).
%
\bibitem{BcXchc6} Chao-Hsi Chang, Yu-Qi Chen, Guo-Li Wang, Hong-Shi Zong,
  {\it Phys. Rev.} {\bf D 65}
014017 (2002).

\bibitem{BcXchc7} P. Colangelo, F. De Fazio, T.N. Pham,  {\it Phys. Rev.}
{\bf D 69} 054023 (2004).
%
\bibitem{BcXchc8} Chao-Hsi Cahng, {\it XXXVIIth Rencontres de Moriond on QCD
and High Energy Hadronic Interactions},  arXiv:hep-ph/0205112.
%
%
\bibitem {12} M. A. Shifman, A. I. Vainshtein, V. I. Zakharov, Nucl. Phys. {\bf B 147} (1979) 385.
\bibitem {ming} Ming Qiu Huang, Phys. Rev. {\bf D 69} (2004) 114015.
\bibitem {neubert1} M. Neubert, Phys. Rep. {\bf 245} (1994) 259.
\bibitem {kazem} T. M. Aliev, K. Azizi, A. Ozpineci, Eur. Phys. J. {\bf
C 51} (2007) 593.
%
\bibitem {kazemmel}  K. Azizi, M. Bayar, Phys. Rev. {\bf D 78}, 054011 (2008).
\bibitem{20} B. L. Ioffe, Prog. Part. Nucl. Phys.
{\bf 56} (2006) 232.
%
\bibitem{16} C. Amsler \textit{et al.,} Particle Data Group, {\it Phys. Lett.} {\bf B 667}
1 (2008).
%
 \bibitem{huang} T. Huang, C. W. Luo, Phys. Rev. {\bf D 50} (1994) 5775.
  %
\bibitem{dai} Y. B. Dai, C. S. Huang, C. Liu, and S. L. Zhu, Phys.
Rev. {\bf D 68}, 114011 (2003).
%
\end{thebibliography}
\end{document}